\documentclass[aps,pre,preprint,groupedaddress]{revtex4}   
\usepackage{graphicx}
\usepackage{bm}
\usepackage{setspace}                         
\usepackage{amsmath}
\usepackage{longtable}
\newcommand{\kmax}{$k_{max}$}
\newcommand{\jmax}{$j_{max}$}
\newcommand{\cth}{$c_{th}$}
\newcommand{\mth}{$m_{th}$}
\newcommand{\rl}{recurrence length}
\newcommand{\lmch}{$r^{m}_{ch}$}
\newcommand{\dagpa}{DAGPA}
\newcommand{\pacz}{BAPA}

\begin{document}

\title{Network of Earthquakes and Recurrences Therein}
\author{Krishna Mohan, T. R.}
\author{Revathi, P. G.}
\affiliation{CSIR Centre for Mathematical Modelling and Computer Simulation (C-MMACS)\\Bangalore 560017, India}
\email{kmohan@cmmacs.ernet.in}
\homepage{http://www.cmmacs.ernet.in/~kmohan}
\date{\today}

\begin{abstract}
We quantify the correlation between earthquakes and use the same to distinguish between relevant causally connected earthquakes. Our correlation metric is a variation on the one introduced by Baiesi and Paczuski (2004). A network of earthquakes is constructed, which is time ordered and with links between the more correlated ones. Data pertaining to the California region has been used in the study. Recurrences to earthquakes are identified employing correlation thresholds to demarcate the most meaningful ones in each cluster. The distribution of recurrence lengths and recurrence times are analyzed subsequently to extract information about the complex dynamics. We find that the unimodal feature of recurrence lengths helps to associate typical rupture lengths with different magnitude earthquakes. The out-degree of the network shows a hub structure rooted on the large magnitude earthquakes. In-degree distribution is seen to be dependent on the density of events in the neighborhood. Power laws are also obtained with recurrence time distribution agreeing with the Omori law. 

\end{abstract}

\maketitle
\section{Introduction}
New ways to analyze earthquake catalogs have emerged in recent times in the wake of an objective approach where the subjective/arbitrary element in the identification of aftershocks, foreshocks etc. have been discarded \cite{Baketal2002, AlCor2004, BaPcz2004, Jdetal2006}.  For example, attempts to identify aftershocks, foreshocks etc. using correlations between events based entirely on data have been suggested \cite{BaPcz2004, Pcz2005, BaPcz2005}. Treating all earthquakes, regardless of foreshock and aftershock characterizations, on the same footing is another approach \cite{Baketal2002, AbeSuz2004, Jdetal2006} which reveals some basic characteristcs that affect them all, and which depends only on magnitude and, perhaps, on finite size effects. That the aftershocks and main shocks are not caused by different relaxation mechanisms and no clear distinction can be made between them is getting accepted by a significant number of researchers \cite{HouJo1997, Kagan2002, Baketal2002, BaPcz2004}.  All this leads to identification of interesting new facets of earthquake dynamics and patterns arising therefrom. 

The clustering of earthquakes in space and time suggests that events that follow in time are, to a certain extent, causally related to the earlier ones. However, restricting the causality connection to the somewhat arbitrary mainshock-aftershock scenario may not be enough. Rather, the causality connection can be extended to a cluster of events that are strongly correlated based on data analysis. We adopt this approach herein and study spatial recurrences by constructing a network of events that are correlated above a certain threshold. While a unique correlation relation may not be possible, we follow Baiesi and Paczuski \cite{BaPcz2004} and base our correlation relation on the two main statistical laws that characterize earthquake data--- the Gutenberg-Richter law \cite{GR1941} and the fractal distribution of earthquake epicenters (see, for example, \cite{Turcot1997}). The former law states that the number, $N$, of earthquakes of magnitude $\ge m$ vary as \begin{equation} \label{Gut} N(m) \sim 10^{-bm} \end{equation} where $b$ is a constant $\approx 1$, but does vary a little with the region and the catalog (see, for example, \cite{BaYiOz}). The latter is based on fractal analysis and is of recent origin, but has been shown to be quite robust through the analysis of various datasets from various regions of the globe (see, for example, \cite{Hirata, Turcot1997}): \begin{equation} \label{fract} N'(l) \propto l^{d_f} \end{equation} where $N'$ is the number of pairs of points separated by a distance $l$ and $d_f$ is the fractal dimension.

If we combine the above two laws, we may state that the average number of events that can occur in the region within a distance $l_{ij}$ separating two events $i$ and $j$ is \begin{equation} \label{nij}  n_{ij} = Kl^{d_{f}}_{ij}10^{-bm_{i}}\Delta m_{i} \end{equation} where $K$ is a constant of proportionality that may be related to the seismic activity of the region \cite{Ba2006}, and $\Delta m_{i}$ is the accuracy in measurement of the magnitude $m_{i}$. Note that the events of the catalog are assumed to be time ordered with $j > i$ and $t_j > t_i$. We may then define a correlation relation between earthquakes as  \begin{equation} \label{eqcor} c_{ij} = \frac{1}{n_{ij}} \end{equation} and note that the correlation between two events is a maximum when $n_{ij}$ is at a minimum.

We construct a growing network by connecting the correlated events together. For this purpose, correlation thresholds, \cth, are defined and only those event pairs with $c_{ij} \ge$ \cth\ are linked together in the network. The basic premise is that all subsequent events in the catalog are recurrences to the selected event, but a pruning is being done based on \cth\ to ensure that only the most correlated are selected. 

Our metric for measuring correlation is similar to the one used by Baiesi and Paczuski \cite{BaPcz2004} (hereafter referred to as \pacz), but, since we are focussing on recurrences rather than aftershocks, we have chosen to omit the time factor $t_{ij} \ \ (= t_j - t_i)$ from the expression occurring in \pacz.  The metric used here ensures that spatially closeby events are more correlated and, also, a higher magnitude event contributes a larger value to correlation than a lower magnitude one at that same location. In particular, our metric does not, unlike in \pacz, leave out events that are separated by large time intervals, as long as they satisfy the correlation relation, Eq.~\ref{eqcor}.

It is to be also noted that, in \pacz, the question that was asked was as to which previous event the present event is an aftershock of. To this end, Eq.~\ref{nij} (with an added factor of $t_{ij}$ in the equation) was evaluated in \pacz\ by holding  $j$ fixed and varying $i$ from the set of all previous events. In such a procedure, $m_i$ as well as $l_{ij}^{d_f}$ keep changing in each evaluation. In our case, we go {\em forwards} in time and ask which event is returning to the same location. Therefore, $i$ and $m_i$ are fixed, $j$ varies between evaluations and,  only $l_{ij}$ keeps varying.  Thus, in our case, each event selects a spatial window proportional to its magnitude and the subsequent events closer to it get more weightage in the computed correlation value. In the case of \pacz, the size of the spatial window depends on, not the magnitude of the event whose associations are being looked for but, the earlier events which are being sampled to identify associations.  Our procedure identifies recurrences whereas aftershocks are identified in \pacz. 

\subsection{The Catalog}
Our results have been based on a study of a seismic event data catalog of California region (for the  rectangular region covering $(120.5^{\circ}\mathrm{W}$--$115.0^{\circ}\mathrm{W})$ longitude and $(32.5^{\circ}\mathrm{N}$--$36.0^{\circ}\mathrm{N})$ latitude), for the period  January, 1984--December, 2002~\cite{califcatalog}. There are many studies pertaining to this region \cite{Grassetal2007, DaPcz2004, Baketal2002, AbeSuz2004, DaPcz2005, AlCorral2003, Ba2006, AbeSuz2004-2, AbeSuz2006, Pcz2005, BaPcz2005,Yangetal2004,Woodard,Corralcomm} and this makes it possible to critically compare results. The catalog is assumed to be homogeneous for this period and is complete for events with $m \ge 2.5$ \cite{AGU2}. The relative location errors are of the order of $100$ m~\cite{BSSA1}. The number of events in the catalog  is $19569$. The $b$ value for the catalog  obtained from a linear fit to the plot of $\log N$ vs. $m$ of Eq.~\ref{Gut} is $0.97$; the linear fit was good in the range $2.5 < m < 6$.

Sub-catalogs with higher magnitude thresholds such as $m_{th} = 3$, $3.5$ and $4.0$ and  subsets  for  shorter periods  were also employed to check the dependence  on $m$  and on the number of earthquakes. 

\section{Methodology}

The absolute value of $K$ is not critical for this study since only the relative correlation values matter in separating the most correlated from the less correlated. We have chosen to set $K = 10^{-9}$ as was done in \pacz. Again, we have taken $d_{f} = 1.6$ as in \pacz\  which is a good representative value given the many estimates available \cite{Sahimietal1993, Turcot1995}. $\Delta m_{i} = 0.1$ as available from the data site \cite{califcatalog}.

\begin{figure}
\caption{The $\log$-$\log$ plot of the probability distribution, $P(c)$,  of the correlation values ($c$) between seismic events, versus $c$. The latter has been binned logarithmically and are between pairs of events in the network satisfying $m \ge$ \mth\  $= 2.5$. The linear fit in this $\log$-$\log$ plot shows that the distribution is a power law with an exponent of $\approx -1.7$. A power law distribution is observed for the other \mth\  values too. The range in $c$ values decreases with increase in \mth.}
\label{corr}
\includegraphics[]{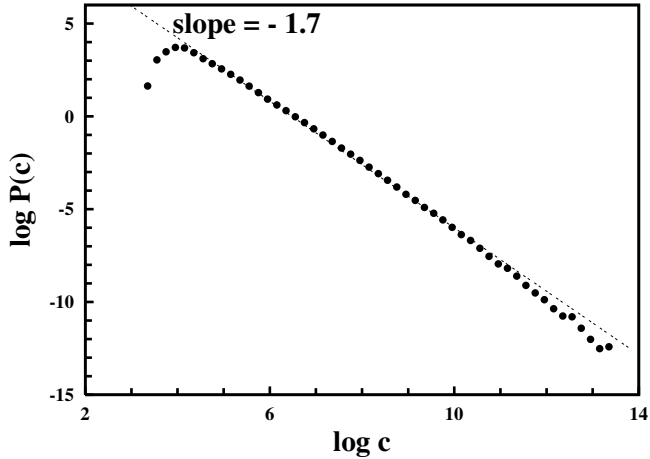}
\end{figure}

The distribution of correlation values between all $i$-$j$ pairs in the data set shows a scale free distribution. This is evident in the $\log$-$\log$ plot of $P(c)$ vs $c$ (Fig.~\ref{corr}), where $c$ is the set of binned correlation values and $P(c)$ the probability values of those bins. The linear regime in this plot ranges over 8 orders of magnitude. The slope of the linear region is   $\approx -1.7$ for the catalog subset with \mth\ $= 2.5$, and (not shown here) $-1.6$ for \mth\ $= 4.0$. The broad distribution implies that, typically, an event $i$ has a few events $j$ which are strongly correlated to it and a larger number of $j$'s which are weakly correlated to it. By specifying correlation thresholds \cth, we may select only those $i$-$j$ pairs from the data set that have $c_{ij}$ $\geq $ \cth. Outgoing links are drawn from each event to another if $c_{ij} \ge$ \cth. Clusters result, where each cluster consists of an event and its  highly correlated recurrences. A cluster can have as few as  two connected events. There may also be isolated events with no clusters around them. By varying \cth, networks of different levels of density are generated.

Since many of the commonly observed patterns in seismic events data have magnitude dependence, networks were also generated by selecting only those events with magnitudes greater than a threshold value, \mth. In this case too, correlations are calculated between the events of the subset of the data points with $m > m_{th}$ and, subsequently, an appropriate \cth\ imposed for generating networks with links between the qualifying pairs of events. 

\begin{figure}
\caption{Network constructed by three different procedures are presented for comparison sake.  Fig.~\ref{network}(a) has been constructed using the procedure advised in \dagpa, Fig.~\ref{network}(b) has been constructed using the procedure of \pacz\ and,  finally, Fig.~\ref{network}(c) employs our own procedure. Same data set has been employed for all the three panels and the epicenter of Hector Mine earthquake has been selected as the originating node of aftershocks/recurrences. Procedure of \pacz\ is restrictive in time while the procedure of \dagpa\ is minimalistic and selcts an event only if it is closer than the previously selected event. Our procedure selects the full set of recurrences with $c_{ij} \ge c_{th}$.}
\label{network}
\includegraphics[]{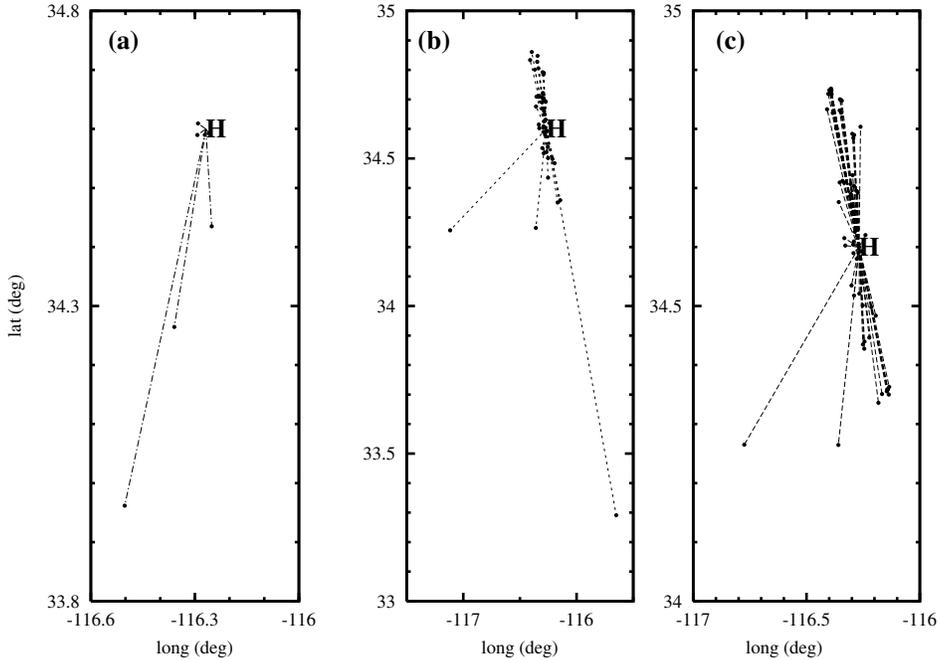}
\end{figure}

We have operated in the range of  $10^{7} \leq c_{th} \leq 10^{10}$. Higher thresholds result in too sparse a network.  While $c_{ij} < 10^7$ was also obtained with many $i$-$j$ pairs in the data set, we have verified that lower \cth\ values do  not yield any new information while at once  making the network too dense and unwieldy. In Fig.~\ref{network}, we present a part of the network for visualization purposes. Three panels have been provided for ready comparison of the differences in the networks generated by three different procedures, viz. by a procedure developed by Davidsen, Grassberger and Paczuski \cite{Jdetal2006} (hereafter referred to as \dagpa; see Section \ref{discuss} for further discussion on this procedure) (Fig.~\ref{network}(a)), by the procedure developed in \pacz\ (Fig.~\ref{network}(b)), and by our own procedure (Fig.~\ref{network}(c)). We have presented in the three panels the data pertaining to the Hector Mine earthquake and its aftershocks/recurrences, as perceived in the three procedures. One can see that our procedure picks up far more events since no time window applies to it. Also, the procedure of \dagpa, since it picks up only the record breaking events has minimal links.

\section{Results}

\subsection{Degree distribution}

The out-degree of  a node in these networks describes the number of its recurrences and the in-degree describes the number of events which count the selected node as a recurrence to themselves.

\begin{figure}
\caption{The linear variation in the $\log$-$\log$ plot of maximum out-degree $(k_{max})$  with \cth\ (Fig.~\ref{degree_distb}(a)) shows that \kmax\ decreases as a power law as \cth\ increases, with an accelerated fall off after \cth\ $= 10^9$. The different graphs are for different \mth\ values and it is seen that the behavior is similar with the  graphs for different \mth\ values being parallel.  The maximum in-degree $(j_{max})$ (Fig.~\ref{degree_distb}(b)) is seen to be couple of magnitudes lower than \kmax\ (cf. Fig.~\ref{degree_distb}(a)). \jmax\ also decreases with \cth\ as a power law, as with \kmax. However, it also becomes magnitude independent for larger \cth\ values as is seen in the figure where all the graphs for different \mth\ values converge to a common value at \cth\ $=10^{10}$. The general out degree ($k$) distribution is similar with a power law observed (Fig.~\ref{degree_distb}(c)) there too; the plots are for a fixed \mth\ ($= 2.5$) and for different \cth\ values (shown by the different symbols), and the power law exponent is $\approx -2.0$. In Fig.~\ref{degree_distb}(d), we see that the in-degree ($j$) distribution, on the other hand, has a plateau for the intermediate $j$ values indicating a uniform distribution on an average.} 
\label{degree_distb}
\includegraphics[scale=1.3]{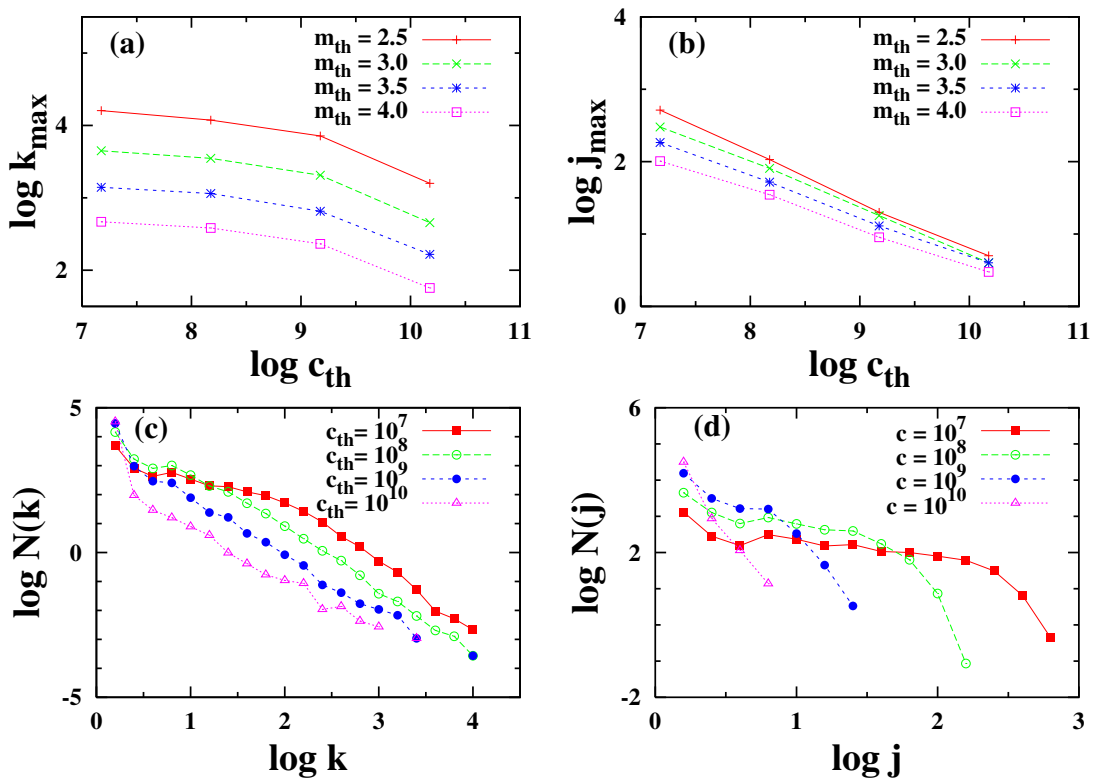}
\end{figure}

Generally speaking, the maximum out-degree (\kmax) obtained in the network (Fig.~\ref{degree_distb}(a)) is at least an order of magnitude higher than the maximum in-degree (\jmax) obtained (Fig.~\ref{degree_distb}(b)), for all \mth\ and \cth\ values. It is clear that large earthquakes will tend to have more events associated with them through higher correlation values since Eq.~\ref{eqcor} is positively correlated with magnitude. On the other hand, lower magnitude events require small $l_{ij}$'s to obtain larger correlation values, i.e. the density of points become important for such events. Since \jmax\ has, consistently, lower values compared to \kmax\ values, this suggests that \jmax\ is dependent on $l_{ij}$ and, thus, the density of points. Given that the event distribution is a power law with respect to magnitude (Eq.~\ref{Gut}) as well as correlation (Fig.~\ref{corr}), the above pattern is sustained with all \mth\ and \cth\ values. To summarize, a hub structure is present as far as out-degree is concerned while it is absent with the in-degree distribution.

As \cth\  is increased and the network becomes sparser, \jmax\ and \kmax\ fall off as  power laws, for all \mth\ values. \jmax\ falls off faster and, at $c_{th} = 10^{10}$, the difference with the out-degree value is about two orders of magnitude. This again confirms the density dependence of the in-degree distribution.  Furthermore, at this higher \cth\ value, \jmax\ becomes magnitude independent as well, and all the graphs for the different \mth\ values converge to the same \jmax.  Similar plots for \kmax\ remain parallel, while falling off, for the different \mth\  values. This indicates that the hub structure survives even when the network becomes sparser. This is understandable since larger magnitude events dominate \kmax\ distribution and they populate the sparse networks produced by the higher \cth\ values. Sparsity induced by higher \mth\ or \cth\ thresholds does affect the rate of fall-off of \kmax. There is a  region with a a slow fall off ($\sim -0.17$) in the beginning followed by an accelerated fall off ($\sim -0.65$), induced by sparsity. The latter sets in at $c_{th} \approx 10^9$ (Fig.~\ref{degree_distb}(a)). 

The general out-degree distribution, $N_k$ vs. $k$ (Fig.~\ref{degree_distb}(c)), where $N_k$ is the number of nodes with degree $k$, mirrors the above scenario seen with maximum out-degree.  A power law behavior is obtained for all correlation thresholds and the range of the power law is almost over the entire range of out-degree values, i.e. over nearly three decades. The exponent of the power law is $\approx -2.0$. Nevertheless, when $k$ is small, the order of magnitude of $N_k$ is practically the same for all correlation thresholds. Since these small $k$ values cannot be asssociated with larger magnitude events, this indicates that the lower magnitude events  have their $k$ values determined by the density of events in the clusters. 

In the case of the in-degree distribution (Fig.~\ref{degree_distb}(d)), the range of degree strengths is much smaller and gets even smaller with an increase in \cth. This again confirms that sparsity affects in-degree significantly since it depends mostly on the density of  events.  A significant plateau is obtained in the middle of the graph, corresponding to intermediate $j$ values, for most \cth\ values, which can be associated with the smaller magnitude events that abound in the network. This indicates that a uniform distribution of $j$ values, prescribed by the density of events, is the dominant characteristic of the in-degree distribution.
 
\subsection{Distribution of recurrence times} 

An important feature of earthquake data is that aftershocks decay according to the Omori law ~\cite{omori1894, utsu1995} \[ n(t) \sim \frac{K}{(c+t)^p} \] where $c$ and $K$ are constants in time but depend on magnitude $m$ and, $p \approx 1$.  $n$ is the number of aftershocks per unit time. Earthquakes of all magnitudes have aftershocks  which decay according to the Omori law. It is widely recorded in literature that even intermediate magnitude events can have aftershocks that persist upto years~\cite{BaPcz2004}. We investigate this by studying the distribution of {\em recurrence times} (also referred to as `waiting times' in the literature \cite{Baketal2002, AlCorral2003}) in the data.  
\begin{figure}
\caption{Recurrence time distribution for \cth\ $= 10^7$  and for two different \mth\ values (2.5 and 4.0) is shown in $\log$-$\log$ plots. Normalized probabilites have been scaled with respect to bin sizes in the plot. Recurrence times $ < 180$ secs have not been plotted. The plots show a power law variation with an exponent of about $-1$ for most of the range on the x axis.  The range over which the plots have a slope of about $-1$ increases with increasing \mth. This has been verified with different \mth\ values and is also seen here from the two graphs for the different \mth\ values. However, plot also becomes more noisy as \mth\ is increased due to data paucity. Range of the power law  remains approximately the same for different \cth\ values.}
 \label{rectime}
\includegraphics[]{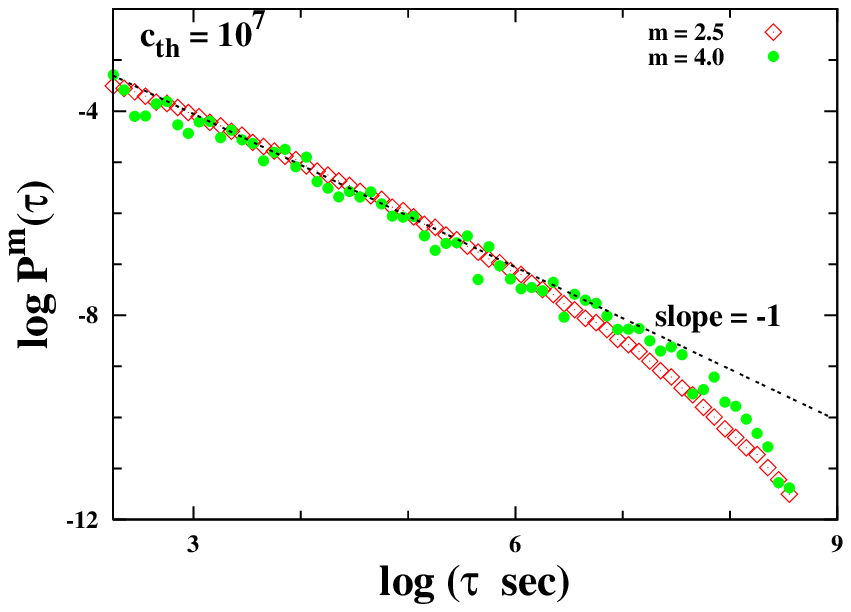}
\end{figure}
We define recurrence time here as the time between successive events in a cluster surrounding an event.  The recurrence times in all the clusters ranges from seconds to tens of years. A $\log$-$\log$ plot of the distribution is given in Fig.~\ref{rectime} where we have left out recurrence times $< 180$ seconds because of the likely error margins. The plot shows that a power law holds for most of the range of recurrence times and, in particular, for about 5--6 decades. Furthermore, the law holds equally well for all $m_{th}$  values as the overlap of the graphs for \mth~$= 2.5$ and \mth~$= 4.0$ shows. The higher recurrence times do show some scatter, but that is understandable given the sparsity of data in that region. There is, however, a small  change in slope for the higher recurrence times. The initial slope of $-1.0$ changes over to $-1.2$ for one or two decades before sparsity of data take over. The range over which the power law holds seems to increase with increase in \mth\ as seen from the graph for \mth\ $= 4.0$ when compared to the graph for \mth\ $= 2.5$. We find it remarkable that a power law holds for the recurrence times just as the Omori law for the aftershocks, which are a subset of the recurrences. In particular, note that the power law is very good up to $\tau \approx 10^7$ s (equivalently $\tau \approx 4$ months). This range in time may be different for different seismic regions. Considering that there is no objective criterion for fixing the time window for identifying aftershocks, we propose that the upper limit of the time window for the same be fixed at the point where the power law fails in the recurrence time plots such as Fig.~\ref{rectime}.

\subsection{Distribution of recurrence lengths}
By recurrence length, we mean the distance from an event to each of the other events in the same cluster. In other words, we are measuring the scatter in the way the subsequent events in the cluster to an event are distributed around the event in question.  While there is no established law like the Omori law for recurrence lengths, studies have indicated that a power law seems to hold for recurrence lengths as well \cite{Grassetal2007}. To investigate this, we use  the great circle distance $r_{ij}$ for computing the distance between two event locations. This is calculated using the  Haversine formula \cite{haversine}: if $(\phi_s, \lambda_s)$ and  $(\phi_f, \lambda_f)$ are the (latitude, longitude) values for the two event locations and, $\Delta \phi = \phi_f - \phi_s$ and  $\Delta \lambda = \lambda_f - \lambda_s$, then \[ r_{ij} = R \Delta \sigma\] where $\Delta \sigma$, the  spherical distance, or central angle, between the two event locations is \[ \Delta \sigma = 2 \arcsin \left( \sqrt{\sin^2 \left(\frac{\Delta \phi}{2}\right) + \cos(\phi_s) \cos(\phi_f) \sin^2 \left(\frac{\Delta \lambda}{2}\right)} \right). \]
$R$, the radius of the earth,  is taken as $6367~ km$.

\begin{figure}
\caption{The recurrent length distribution for \cth\ $= 10^{7}$ is shown in Fig.~\ref{RLplots}a and for \cth\ $= 10^{9}$ is shown in Fig.~\ref{RLplots}b. Both are $\log$-$\log$ plots. Again, normalized probabilites have been scaled with respect to bin sizes in the plot.  The different symbols corresspond to different \mth\ values. All the plots are seen to be unimodal and peaks at characteristic $r$ values, denoted as $r^{m}_{c_{h}}$, that  depend only on  \mth, being larger for higher values of \mth. The  exponent of the power law for the part before the peak is independent of  $m_{th}$ but depends on \cth\ and goes from  $0.45$ for \cth\ $= 10^7$ to $0.55$ for $c_{th} = 10^9$. The exponent of the power law  for the part after the peak is also independent of $m_{th}$, but again depends on \cth; it goes from  $-1.2$ for \cth\ $= 10^7$ to $-1.08$  for \cth\ $= 10^9$. There is, understandably, some scatter for the higher $r$ values because of data paucity. As \cth\ is increased, the peaks shift to lower and lower $r$ values until, for \cth\ $= 10^{10}$, it disappears altogether when it falls below 100 m; note that we have left out $r$ values $< 100$ m due to accuracy constraints. By rescaling $x$  and $y$ axes suitably, all the graphs can be made to collapse to a single curve as shown in the insets. The rescaling function can be represented as $P_{m}(r) = 10^{-\alpha m_{th}}F(r/10^{\alpha m_{th}})$ where $\alpha = 0.51$.  $\alpha$ is the same for all $c_{th}$ values. If the peaks for a particular \mth\ is plotted against \cth\  (not shown here), we find again a power law with an exponent of $-0.5$ which is the same for all \mth\  values.}
\label{RLplots}
\includegraphics[]{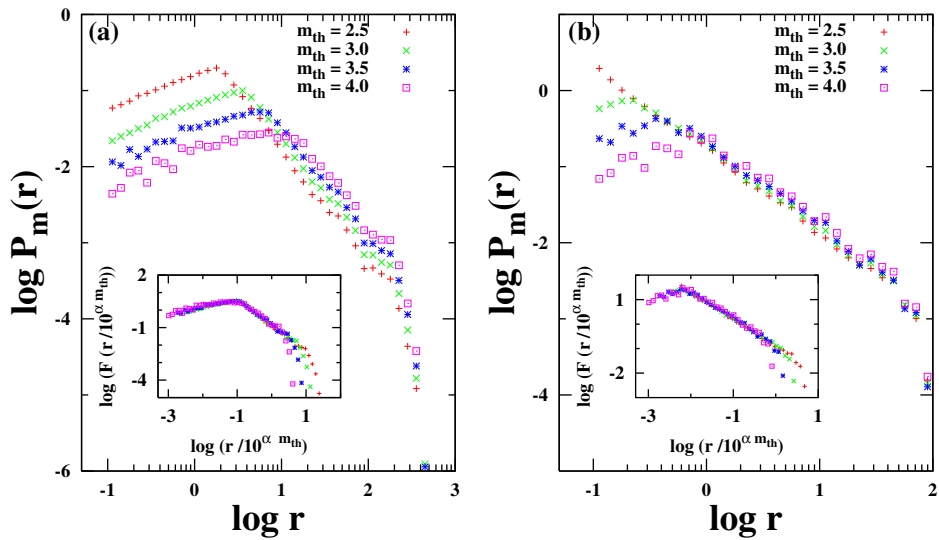}
\end{figure}
The probability distribution of recurrence lengths is unimodal in nature and peaks at a \rl\ ($r^{m}_{ch}$) which is larger for increasing values of $m_{th}$, for a fixed $c_{th}$ (Fig.~\ref{RLplots}).  Note that this is consistent with the usual practice of using larger space-time windows to collect aftershocks of larger magnitude events. Keep in mind that the distribution with lower \mth\ value contains the data points appearing in the distribution with higher \mth\ values. Also, not all the data points appearing with the lower \mth\ value is present with the higher \mth\ value. This points to the presence of a characteristic \lmch\ associated with a particular $m$ value and this clearly is larger for larger $m$ values. Since the rupture is larger with larger earthquakes, we may think of \lmch\ as a typical rupture length associated with events of that magnitude (see also \cite{Grassetal2007}). 

\lmch\ varies with \cth\ as well and the peaks for a fixed \mth\ shifts to lower values as \cth\ is increased. As a result, the earlier part of the graph where it rises to the peak gets progressively shortened as \cth\ increases and the network becomes sparser. Eventually, for \cth\ $= 10^9$, we see only the part of the graph after the peak. Note that we are plotting only pairs of events with $r_{ij} >$ 100 m. As such, the peaks disappear for larger values of $c_{th}$ since the peak values are now $<$ 100 m. The behaviour seen with \lmch\ is completely in accordance with Eq.~\ref{nij}. As \cth\ increases, we are selecting event pairs of larger magnitudes separated by smaller distances. The fact that \lmch\ gets smaller with higher \cth\ suggests also that there is a range of rupture lengths associated with any magnitude, and not a very sharply defined rupture length. Both the rise to the peak value as well as the subsequent fall off are in the form of power laws. The slope of the rise to the peak varies with \cth,  but stays same for different \mth\ values for a fixed \cth. 

While the graphs for different $m_{th}$ are separated before the rise to the peak, they are collapsed onto a single curve after their peaks. The latter falls off as a power law with an exponent $\approx -1.2$ before data paucity forces scatter. The exponent reduces to  $-1.08$ as \cth\  is increased and number of events decreases.  The  plots for the  different values of \mth\   collapse on to a single curve when the $x$ axis is rescaled as $l/10^{\alpha m_{th}}$  and the $y$ axis by  $10^{\alpha m_{th}} P_{m}(r)$ with $\alpha$ = $0.51$. The inset plots in  Fig.~\ref{RLplots} shows the data collapse after the rescaling. Note that the value of $\alpha$  remains the same for all \cth\ values.  The  value of the rescaling parameter $\alpha$ was arrived at by studying the dependence of \lmch\   with \mth.  Since  both \mth\  and \cth\  are arbitrarily chosen values, the robustness of $\alpha$, as the latter  is changed, shows that the \lmch\ values are reliably related across the scale invariant structure of the network. Our studies with other data sets \cite{KM_comp} show that the value of $\alpha$ depends on the seismicity of the region and is not universal.

\section{Discussion and Conclusions}\label{discuss}

Network analysis of earthquake events have grown in strength and many different approaches have been attempted in constructing the networks (see, for example, \cite{AbeSuz2004, BaPcz2004, Jdetal2006}). Nevertheless, some of the conclusions from these different approaches are the same. For example, all are agreed that the out degree distribution is scale free and has a hub structure. However, there are also differences in the conclusions.

In the earlier work of \dagpa, recurrences were studied using a network approach. However, the emphasis was on {\em record breaking} events which were defined as those which come closer to the present one than any earlier. The event that occurred next in time to the present one was taken as the first record and, subsequently, record breaking events were used to build up the set of records (equivalently, recurrences). In particular, the magnitude of a subsequent event was not taken into consideration and the recurrences were taken strictly on spatial closeness criterion. It is possible to construct a  minimal network of recurrences by stipulating that only record breaking events qualify as recurrences. Even though the definition of immediate successor in time to the present event as a recurrence is not satisfactory, nevertheless a minimal network will eventually discard those events when a record breaking event is found.

The data set used in \dagpa\ is almost the same as the one used in this present study and the distribution of recurrence lengths in \dagpa\ turns out to be similar to the one here (Fig.~\ref{RLplots}). This is not surprising because it is a minimal network of recurrences, as compared to the full set of recurrences considered here.  A characteristic length emerges, with the minimal network, at which the distribution peaks which was identified in \dagpa\  to be a typical rupture length associated with that magnitude threshold. The slope of the part after the peak has the same slope as what is obtained in the present study. However, the part before the peak was claimed to have a slope of unity confirming a random distribution of events within a disc of 100 m (the location error bar in the data set) around any event. However, we find a power law behaviour in this part too with a growth slope $\sim 0.5$. In particular, for each \cth, the graphs for different \mth\ have the same slope and are parallel. As \cth\ is increased, this part is shortened and eventually disappears for \cth\ $= 10^{10}$. Since only larger magnitude events are present with these high \cth\ values, this points to the fact that the big events do not cluster at closeby distances. On the other hand, this also implies that small magnitude events predominantly cluster at nearby distances. \dagpa\ reports that no systematic errors have been reported in literature for magnitude dependence of location errors, and argues that the location errors which are $< 100$ m, tend to randomize the distribution for recurrence lengths $< 100$ m.  However, the order we see in the portion before the peak in Fig.~\ref{RLplots}, and similar plots for different \cth\ values, suggests that location errors could be magnitude dependent and should be looked for. It is, of course, to be emphasized that our analysis includes only the correlated earthquakes which brings in the order seen in the portion before the peak.

It is instructive to compare our results for recurrence times with another study on `waiting times'. Bak et al. \cite{Baketal2002} had analyzed data from Southern California and shown that a unified scaling law  exists for waiting times. They had divided the study area into grids of varying sizes and compared the waiting time distribution across the grid. They showed that a scaling law of the form \[ T^{\alpha} P_{S,L}(T) = f\left(TS^{-b}L^{d_f}\right) \] holds where $P$ is the waiting time distribution, $T$ the waiting time, $m = \log S$ and $L$ is the linear size of the grid. The  function $f$ was seen to consist of a constant part and a decaying part, separated by a sharp kink. They concluded that the constant part consisted of the correlated earthquakes and the decaying part with the uncorrelated earthquakes, i.e. independendent earthquakes seeded by the nonzero driving rate of plate tectonics. Several other studies have since appeared  analyzing this unified scaling law which combines Eq.~\ref{Gut}, Eq.~\ref{fract} as well as the `waiting times' \cite{AlCorral2003, AlCor2004, AlCor2006, TalbiYama2009}.

Our studies are based on the same area and practically the same data set. However, we have divided the events into clusters based on correlation values. In other words, we have picked only the correlated earthquakes for analysis. There is, hence, no recurrence times coming from uncorrelated earthquakes. Therefore, there is no kink in the data as seen in Bak et al. \cite{Baketal2002} and our plot has only one slope which agrees qualitatively with the portion of the graph in  Bak et al. \cite{Baketal2002} with correlated earthquakes.

As already mentioned, our definition of the correlation, Eq.~\ref{nij} and Eq.~\ref{eqcor}, differs from \pacz\ in that we have not included $t_{ij}$ in our definition and, also, we evaluate the correlations going forwards in time as against the backward associations in \pacz. The approach developed in \pacz\ is able to address the question of identifying aftershocks in an unbiased manner and without imposing pre-determined space-time windows. On the other hand, we expect the present approach, coupled with the induction of $t_{ij}$, to identify foreshocks, which we intend to pursue in a future work. Finally, we are also proposing that the upper limit of the time window where the power law holds in recurrence time distribution plots such as Fig.~\ref{rectime} could be used as an objective criterion to fix the limit of the aftershock window.
\bibliographystyle{abbrv}
\bibliography{recurrence_compare}

\end{document}